\documentclass[preprint,showpacs,preprintnumbers,amsmath,amssymb]{revtex4-1}
\usepackage{graphicx, latexsym, amssymb, amsmath, color, multirow, mathrsfs, booktabs, ifpdf}
\usepackage[toc,page]{appendix}
\usepackage{dcolumn}
\usepackage{bm}

\def\pN{proton-nucleon\ }
\def\pA{proton-nucleus\ }

\def\hp6{$^6$He+$p$\ }
\def\ph8{$^8$He+$p$\ }
\def\pa4{$p+^4$He\ }
\def\pli6{$p+^6$Li\ }
\def\phe68{$^{6,8}$He+$p$\ }

\begin{document}
\title{Elastic proton scattering at intermediate energies as a probe of 
 the $^{6,8}$He nuclear matter densities}
\author{Le Xuan Chung$^{1}$}
\author{Oleg A. Kiselev$^2$}
\author{Dao T. Khoa$^1$}
\author{Peter Egelhof$^2$}
\affiliation{\centerline{$^1$Institute for Nuclear Science and Technique, VINATOM} 
 \centerline{179 Hoang Quoc Viet Road, Cau Giay, Hanoi, Vietnam.} 
 \centerline{$^2$GSI Helmholtzzentrum f\"ur Schwerionenforschung, 
64291 Darmstadt, Germany.}} 
\date{\today}
\begin{abstract}
The Glauber model analysis of the elastic $^{6,8}$He+$p$ scattering data 
at energies around 700 MeV/nucleon, measured in two separate experiments 
at GSI-Darmstadt, has been done using several phenomenological parametrizations
of the nuclear matter density.  By taking into account the new data points 
measured at the high momentum transfer, the nuclear matter radii of $^{6,8}$He 
were accurately determined from the Glauber model analysis of the data, with the 
spin-orbital interaction explicitly taken into account. The well-known geometry for 
the core and dineutron halo has been used with the new parametrizations of the 
$^{6}$He density to extract the detailed information on the structure of $^{6}$He in 
terms of the core and dineutron halo radii. An enhanced sensitivity of the data 
measured at the high momentum transfer to the core part of the $^{6,8}$He densities 
has been found. 
\end{abstract}
 \maketitle

\section{Introduction}
As $^{6,8}$He beams with high energy resolution and intensity became available 
at different radioactive ion beam facilities around the world, these unstable 
helium isotopes are among the most studied light neutron-rich nuclei. Ever 
since the pioneering measurement of the interaction cross section in the 
late eighties \cite{Tan96,Tan88} which lead to the discovery of the extended 
density distribution of the valence halo neutrons in $^{6,8}$He, many recent 
experimental efforts are still focused on a more precise determination 
of the nuclear radii and radial shape of these nuclei by different methods 
\cite{Tan13}. The elastic proton scattering in inverse kinematics at 
energies around 700 MeV has been proven to be an accurate method to obtain 
information on the nuclear density distributions of the halo nuclei under 
study \cite{Al97,Eg01,Dob06,Ili12}. The first experiment on the (inverse 
kinematics) elastic proton scattering on $^{6,8}$He at energies around 700 MeV 
has been performed at GSI Darmstadt using the hydrogen-filled ionization chamber 
IKAR which simultaneously served as a gas target and a detector for the recoil 
protons \cite{Neu02}, and the measured elastic scattering data were analyzed within 
the Glauber model \cite{Al78} to deduce the matter radii and radial shape 
of the nuclear density distributions of these nuclei \cite{Al02}. These same
data were also studied in a Glauber few-body calculation of the elastic \phe68 scattering 
\cite{Al98}, where the few-body degrees of freedom were treated explicitly. 
We note that the first measurement \cite{Neu02} has covered only the region of low 
momentum transfer because the IKAR active target was limited to the detection 
of recoil ions close to $\theta_{\rm lab}\approx 90^\circ$. Recently, a new 
experimental setup has been designed to study proton induced reactions on the exotic 
nuclei in inverse kinematics using a liquid hydrogen target adapted to obtain 
low-background data \cite{Kis11}. The new setup was successfully used to measure 
the elastic \phe68 scattering at energies around 700 MeV/nucleon, and the measured 
cross section has been extended to the region of higher momentum transfer as 
compared to the previous experiments.  
\begin{figure}[t]
\centering
\includegraphics[angle=0,scale=0.42]{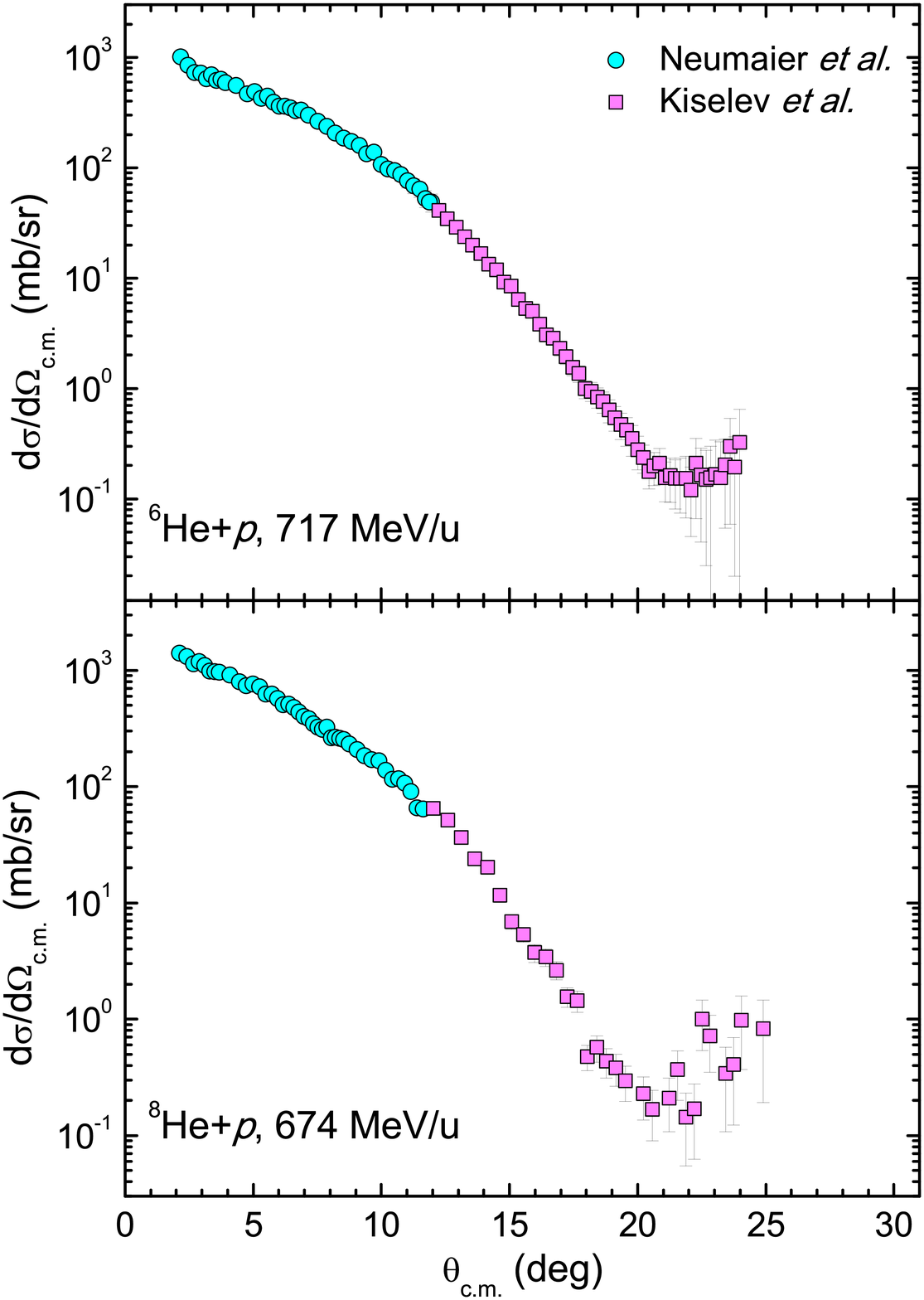} 
\caption{(Color online) Elastic \phe68 scattering data at the energies around 
700 MeV/nucleon measured by Neumaier {\it et al.} \cite{Neu02} and by Kiselev 
{\it et al.} \cite{Kis11} at the low and high momentum transfers, respectively.} 
\label{f1}
\end{figure}
We note that the considered \phe68 data \cite{Neu02,Kis11} were originally 
deduced in terms of the scattering cross section versus the 4-momentum 
transfer squared ($d\sigma/dt$). For a comparison between the results of different 
models, it is more convenient to use the elastic \phe68 scattering cross 
section versus the scattering angle ($d\sigma/d\Omega$) in the 
center-of-momentum (c.m.) frame. The two cross sections are expressed 
through each other by the relativistic kinematics \cite{Rela}   
\begin{eqnarray}
\cos\theta_{\rm c.m.}&=&1+\frac{t}{2k^2} \nonumber \\
 \Rightarrow {\frac{d\sigma(\theta_{\rm c.m.})}{d\Omega}}&=&
 \frac{k^2}{\pi}\frac{d\sigma(t)}{dt}.  \label{eq1}
\end{eqnarray}
Here $-t$ and $k$ are the 4-momentum transfer squared and c.m. momentum, 
respectively. In terms of the scattering angles, the new data points measured 
at high momentum transfer \cite{Kis11} have reached the angular region around 
the first diffractive minimum  (see Fig.~\ref{f1}), and should be, therefore, 
more sensitive to the inner radial part of the ground-state densities of the 
$^{6,8}$He nuclei. The (first) diffractive minimum was observed at 
$\theta_{\rm c.m.}\approx 20^\circ\div 25^\circ$ in the $^6$He+$p$ and 
$^8$He+$p$ data taken at high momentum transfer.   

In the present work, the elastic $^{6,8}$He+$p$ scattering data under study 
have been analyzed within the Glauber multiple-scattering model (GMSM) using the
same phenomenological parametrizations of the matter densities of $^{6,8}$He 
as those used in the earlier GMSM analyses of the GSI data \cite{Al02}. Because 
the two measurements of Refs.~\cite{Neu02, Kis11} were done practically at  
the same energy, it is possible to combine these two data sets in the present 
GMSM analysis. 

\section{Glauber multiple-scattering model}
The basic formalism of the GMSM has been given in details in Ref.~\cite{Al78}.
The GMSM was successfully used in Refs.~\cite{Al02,Al97,Eg01} to analyze 
the elastic $^{6,8}$He+$p$ scattering data measured at low momentum transfer, 
and to deduce the nuclear matter density distributions for these nuclei. However, 
the GMSM calculations in Refs.~\cite{Al02,Al97} were performed without taking 
into account the spin-orbital (s/o) interaction, because the s/o effects were 
known to be negligible at the most forward angles (low momentum transfer) 
\cite{Al78}. Given the new data measured at high momentum transfer which cover 
the first diffractive minimum, the s/o effects should be significant and 
can no more be neglected (see, e.g., Fig.~23 of Ref.~\cite{Al78}). In the 
present work, the formalism of the GMSM that takes into account the s/o 
interaction has been used in the analysis of the new $^{6,8}$He+$p$ data.

The \pA elastic scattering cross section is determined from the elastic
scattering amplitude $F_{\rm el}$ as
\begin{equation}
{\frac{d\sigma}{d\Omega}}_{\rm c.m.}=|F_{\rm el}(\bm q)|^2. 
\end{equation}
In general, the scattering amplitude can be written as \cite{Al02,Glau70}
\begin{equation}
 F_{\rm el}(\bm q) = \frac{ik}{2\pi}\int e^{i\bm{qb}}\left\{1- 
\prod\limits_{j = 1}^A \left[1-\gamma_{pN}(\bm b - \bm s_j) \right]\right\}
\rho_A(\bm r_1, \bm r_2, ..., \bm r_A) \prod\limits_{j = 1}^A d^3r_jd^2b. 
 \label{Amp}
\end{equation}
For the light He nuclei, the effect of the center-of-mass motion 
should be significant. To effectively remove the spurious c.m. motion, 
the \pA scattering amplitude (\ref{Amp}) is multiplied by a correction 
factor $H_{\rm c.m.}(\bm q)$ 
\begin{equation}
H_{\rm c.m.}(\bm q)=\exp\left[\frac{{\bm q}^2R_{\rm m}^2}{6(A - 1)}\right],
\end{equation}
where $R_{\rm m}$ is the root-mean-square matter radius. Such a procedure 
is exact for the nucleon distributions of Gaussian form, and also expected 
to be accurate for the cases of non-Gaussian distributions \cite{Al78}.  

Like the previous GMSM studies \cite{Al02}, we have used in the present 
analysis several density models that divide explicitly the nuclear 
many-body density $\rho_A(\bm r_1, \bm r_2, ..., \bm r_A)$ into the core 
$\rho_c(r)$ and halo $\rho_h(r)$ parts, so that
\begin{equation}
\rho_A(\bm r_1, \bm r_2, ...,\bm r_A)=\prod\limits_{i=1}^4{\rho_{\rm core}(r_i)}
\prod\limits_{j=5}^A{\rho_{\rm halo}(r_j)}. \label{den_ch}
\end{equation}
The representation (\ref{den_ch}) of the many-body density neglects the 
correlations between the nucleon locations, with a constraint that the 
positions of the core and halo nucleons are treated explicitly. In other   
cases, the nuclear many-body density has been simply assumed as  
a product of the one-body matter densities $\rho_{\rm m}(r)$
\begin{equation}
\rho_A(\bm r_1,\bm r_2,...,\bm r_A)=\prod\limits_{j=1}^A {\rho_{\rm m}(r_j)}.
 \label{den_m} 
\end{equation}

In the notations of Eq.~(\ref{Amp}), $\bm b$ is the impact parameter, $\bm q$ is 
the momentum transfer, and $A$ is the nuclear mass number. The proton-nucleon 
($pN$) profile function $\gamma_{pN}$ is determined from the amplitude $f_{pN}$
of the free $pN$ scattering as
\begin{equation}
\gamma_{pN}(\bm b)=\frac{1}{2\pi ik}\int\exp(-i\bm{qb})f_{pN}(\bm q)d^2q.
\end{equation}
In difference from Refs.~\cite{Neu02,Al02}, the present GMSM calculation 
adopts the following parametrization of $f_{pN}$ that takes into account 
also the s/o interaction  
\begin{equation}
f_{pN}(\bm q)=f^{\rm c}_{pN}(\bm q)+ \bm\sigma(\hat{\bm b}\times\hat{\bm k}) 
f^{\rm s}_{pN}(\bm q),\ {\rm with}\ \hat{\bm b}=\bm b/b,\ \hat{\bm k}=\bm k/k.
\end{equation}
Here, $f^{\rm c}_{pN}(\bm q)$ and $f^{\rm s}_{pN}(\bm q)$ are the central and 
s/o parts of the $pN$ scattering amplitude, $\bm\sigma$ is the Pauli spin 
operator. We have parametrized the $f^{\rm s}_{pN}$ amplitude in the same 
way as in Refs.~\cite{Aug76,Ray79}, taking into account explicitly the isotopic 
difference between the total neutron and proton cross sections. Thus, one has 
\begin{eqnarray}
 f^{\rm c}_{pN}(\bm q) &= &\frac{k\sigma_{pN}}{4\pi}(\varepsilon_{pN}+i)
\exp\left(-\frac{{\bm q}^2\beta_{pN}}{2}\right),\ N=p,n \nonumber \\
 f^{\rm s}_{pN}(\bm q) &=& \frac{k\sigma_{pN}}{4\pi}\sqrt{\frac{{\bm q}^2}{4M^2}}
 (i\alpha _{\rm s}-1)D_{\rm s}\exp\left(-\frac{{\bm q}^2\beta_{\rm s}}{2}\right). 
 \label{NAmp}  
\end{eqnarray}
Here $\sigma_{pN}$ is the total $pN$ cross section, parameters $\varepsilon_{pN}$ 
and $\alpha_{\rm s}$ give the ratios of the real and imaginary strengths, 
$\beta_{pN}$ and $\beta_{\rm s}$ are the slope parameters, $D_{\rm s}$ is the 
relative strength of the s/o amplitude, and $M$ is the nucleon mass.  

\begin{figure}[b]
\centering
\includegraphics[angle=0,scale=0.45]{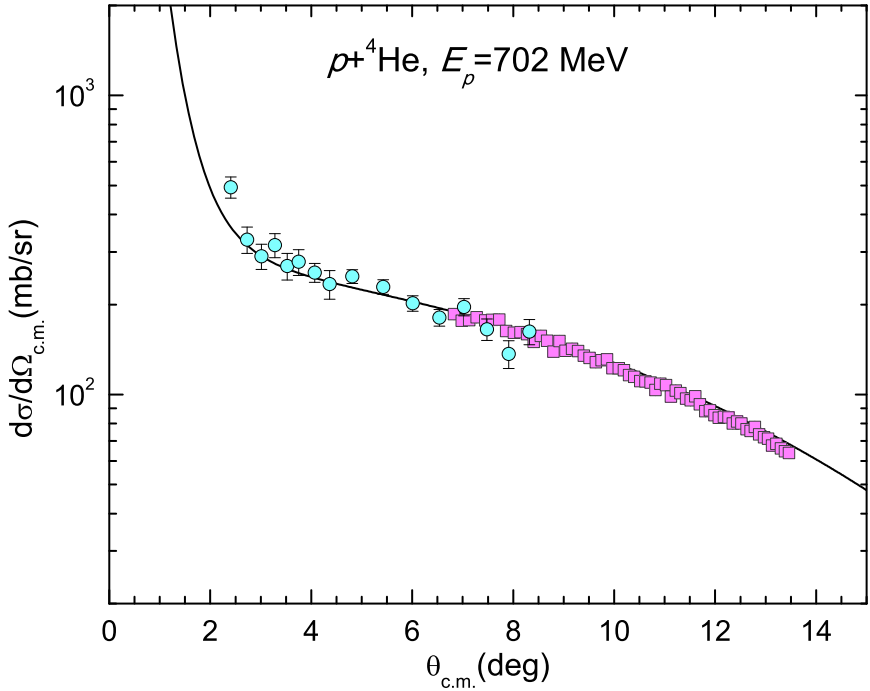}
\caption{(Color online) Elastic \pa4 scattering data measured at proton energies around 
700 MeV (circles \cite{Neu02} and squares \cite{Gre89}) in comparison with the 
elastic scattering cross section given by the GMSM calculation (solid line), taking 
into account the s/o term and using a Gaussian density for $^4$He that gives 
$R_{\rm m}=1.47$ fm.}  \label{He4}
\end{figure}

In the present work we have assumed the same parameters for the central amplitude 
$f^{\rm c}_{pN}$ as those used earlier in Ref.~\cite{Al02}, except the slope 
parameters $\beta_{pN}$ that were fine tuned to obtain the best description of
the elastic \pa4 data at $E_p\approx 700$ MeV \cite{Neu02,Gre89} in the GMSM 
calculation that takes into account the s/o interaction explicitly. The reason 
is that the $\beta_{pN}$ values used in Ref.~\cite{Al02} were adjusted to the 
best GMSM description of the same \pa4 data without taking into account the 
s/o term. Thus, $\beta_{pN}$ and parameters of the s/o term have been readjusted 
in the present work to the best description of the elastic \pa4 data at 
700 MeV, as shown in Fig.~\ref{He4}. All the parameters used in the present 
GMSM calculation are given in Table~\ref{t1}, with the newly obtained 
$\beta_{pN},\ D_{\rm s},\ \beta_{\rm s}$, and $\alpha_{\rm s}$ 
values being quite close to those suggested earlier in Ref.~\cite{Ray79}. 
The GMSM results shown in Fig.~\ref{He4} agree also with the fully quantal 
optical model results given by the complex \pa4 optical potential 
obtained from the folding model calculation \cite{Kho02} using the same Gaussian 
density for $^4$He and finite-range t-matrix interaction by Franey and Love 
\cite{Fra85}. This validates the parameters chosen for the present
GMSM calculation.   
\begin{table}[t]
\centering
\caption{Parameters of the central and s/o scattering amplitudes (\ref{NAmp}) 
used in the present GMSM analysis of the elastic \phe68 scattering.} \label{t1} 
\vspace{0.5 cm}
\begin{tabular}{|c|c|c|c|c|c|c|c|c|c|c|} \hline
System & $E_p$ & $\sigma_{pp}$ & $\varepsilon_{pp}$ & $\sigma_{pn}$ & 
$\varepsilon_{pn}$ & $\beta_{pp}$ & $\beta_{pn}$ & $D_s$ & $\alpha_s$ &
$\beta_s$ \\ 
 & (MeV) & (mb) &  & (mb) &	 & (fm$^2$) & (fm$^2$) & & & (fm$^2$) \\ \hline
\ph8	& 674 & 41.9 & 0.129 & 37.4 & -0.283 & 0.20 & 0.24 & 0.284 & 13.50 & 0.522 \\ \hline
\hp6	& 717 & 44.6 & 0.069 & 37.7 &-0.307 & 0.20 & 0.24 & 0.284 &  13.50 & 0.522 \\ \hline
\end{tabular}
\end{table}

Using the profile function $\gamma_{pN}$ determined by the $pN$ scattering 
amplitude (\ref{NAmp}) and treating the Coulomb term in the standard way 
\cite{Al78,Glau70}, the \pA elastic scattering amplitude can be written as 
\cite{Fal78,Al78}
\begin{equation}
F^2_{\rm{el}}(q)=|F_{\rm {Coul}}(q)+F_{\rm c}(q)|^2+|F_{\rm s}(q)|^2, \label{CS}
\end{equation}
where $F_{\rm c}$ and $F_{\rm s}$ are the central and s/o \pA amplitudes 
\cite{Al78,Fal78}
\begin{eqnarray}
F_{\rm c}(q) &= & ikH_{\rm {CM}}(q)\int[1-G_{\rm c}(b)]\exp[i\chi_{\rm {Coul}}(b)]
 J_0(qb)bdb,  \nonumber \\
F_{\rm s}(q) &= &-kH_{\rm {CM}}(q)\int G_{\rm s}(b)\exp[i\chi_{\rm {Coul}}(b)]
 J_1(qb)bdb. \label{e2}
\end{eqnarray}
The $G$ functions contain explicitly the central and s/o contributions as    
\begin{eqnarray}
G_{\rm c}(b)=\frac{1}{2}\left\{\prod\limits_{j = 1}^A[1-\Gamma^{\rm c}_j(b)+
\Gamma^{\rm s}_j(b)]+\prod\limits_{j = 1}^A[1-\Gamma^{\rm c}_j(b)-
\Gamma^{\rm s}_j(b)]\right\}, \nonumber \\
G_{\rm s}(b)=\frac{1}{2}\left\{\prod\limits_{j = 1}^A[1-\Gamma^{\rm c}_j(b)+ 
\Gamma^{\rm s}_j(b)]-\prod\limits_{j = 1}^A[1-\Gamma^{\rm c}_j(b)- 
\Gamma^{\rm s}_j(b)]\right\}. \label{gam}
\end{eqnarray}
The nucleon profile functions $\Gamma^{\rm c}$ and $\Gamma^{\rm s}$ are 
determined as
\begin{eqnarray}
\Gamma^{\rm c}_j(b) &=& -\frac{i}{k}\int f^{\rm c}_{pN}(q)S_j(q)J_0(qb)qdq, 
 \nonumber \\
\Gamma^{\rm s}_j (b) &=& -\frac{1}{k}\int f^{\rm s}_{pN}(q)S_j(q)J_1(qb)qdq. 
 \label{e3} \end{eqnarray}
Here $J_{0,1}(x)$ are the zero-th and first-order Bessel functions. 
$F_{\rm {Coul}}(q)$ and $\chi_{\rm {Coul}}(b)$ are the Coulomb amplitude 
and phase, respectively \cite{Glau70}. In difference from the earlier GMSM
calculations \cite{Al02,Dob06,Ili12}, the Sommerfeld parameter (used to determine 
the Coulomb term) obtained with the relativistic kinematics has been used in the
present work. The form factor $S_j(q)$ is determined by the Fourier transform 
of the single-particle density as
\begin{equation}
S_j(q)=\frac{1}{H_{\rm {CM}}(q)}\int\exp(i\bm{qr})\rho_j(r)d^3r.
\end{equation}
When one writes explicitly the products in Eq.~(\ref{gam}) in terms of the
nucleon profile functions, the \pA scattering amplitude becomes a multiple 
scattering series \cite{Al78}. When the s/o term is neglected, the amplitude 
(\ref{CS}) is simplified to that used in the earlier GMSM calculation 
that did not take into account the s/o interaction \cite{Al02,Dob06,Ili12}. 
\section{Nuclear densities}
\subsection{Parametrization of the nuclear matter distribution}
In addition to the \pN scattering amplitudes, the nuclear matter density 
distribution is a vital input of the Glauber model calculation. Like in the 
previous studies \cite{Al02,Glau70}, the nucleon point-density has been 
parametrized in the following phenomenological forms. 

\subsubsection{The symmetrized Fermi (SF) density}
The SF density distribution is parametrized \cite{Al02} as  
\begin{equation}
\rho_{\rm m}(r)=\frac{3}{4\pi R_0^3}\left[4+\left(\frac{\pi a}{R_0}\right)^2
\right]^{-1}\sinh\left(\frac{R_0}{a}\right)\left[\cosh\left(\frac{R_0}{a}\right)
+\cosh\left(\frac{r}{a}\right)\right]^{-1}, \label{SF} 
\end{equation}
where $R_0$ is the half-density radius (at which the density becomes twice 
smaller than at the origin) and $a$ is the diffuseness parameter. The 
corresponding root-mean-square (rms) matter radius $R_{\rm m}$ is given by
\begin{equation}
R_{\rm m}=\langle r^2\rangle^{1/2}=\left(\frac{3}{5}\right)^{1/2}R_0
\left[1+\frac{7}{3}\left(\frac{\pi a}{R_0}\right)^2\right]^{1/2}.
\end{equation}

\subsubsection{The Gaussian-Halo (GH) density}
The GH density distribution is determined \cite{Al02} as a function of the rms 
radius $R_{\rm m}$ as 
\begin{equation}
\rho_{\rm m}(r)=\left(\frac{3}{2\pi R^2_{\rm m}}\right)^{3/2}[1+
 \alpha\varphi(r)]\exp\left(-\frac{3r^2}{2R^2_{\rm m}}\right) 
\end{equation}
 with
\begin{equation}
\varphi(r)=\frac{3}{4}\left[5-10\left(\frac{r}{R_{\rm m}}\right)^2+
3\left(\frac{r}{R_{\rm m}} \right)^4 \right]. \label{GH}
\end{equation}

\subsubsection{The Woods-Saxon (WS) density}
The WS density has been used by Glauber in his pioneering work \cite{Glau70} 
\begin{equation}
\rho_{\rm m}(r)=\frac{C}{1+\exp\left(\displaystyle\frac{r-R_0}{a}\right)}, 
 \label{WS} 
\end{equation}
where $R_0$ and $a$ are the same parameters as those used in Eq.~(\ref{SF}),
and $C$ is normalized such that (\ref{WS}) is the nucleon point-density.

\subsubsection{The Gaussian-Gaussian (GG) density}
In the GG parametrization the locations of the core and halo nucleons are 
treated explicitly, with both the core and halo densities assumed to be 
in the Gaussian form \cite{Al02}
\begin{equation}
\rho_{\rm core(halo)}(r)=\left({\frac{3}{2\pi R_{\rm c(h)}^2}}\right)^{3/2}
\exp\left(-\frac{3r^2}{2R_{\rm c(h)}^2}\right). \label{GG}
\end{equation}

\subsubsection{The Gaussian-Oscillator (GO) density}
The GO density has the same Gaussian core as in the GG case, but the halo 
distribution is parametrized using the $p$-shell harmonic oscillator 
wave function \cite{Al02}
\begin{eqnarray}
\rho_{\rm core}(r)& = &\left(\frac{3}{2\pi R_{\rm c}^2}\right)^{3/2}
 \exp\left(-\frac{3r^2}{2R_{\rm c}^2}\right) \nonumber \\
\rho_{\rm halo}(r)& = &\frac{5}{3}\left(\frac{5}{2\pi R_{\rm h}^2}\right)^{3/2} 
\left(\frac{r}{R_{\rm h}}\right)^2\exp\left(-\frac{5r^2}{2R_{\rm h}^2}\right).
 \label{GO} \end{eqnarray}

Because the GG and GO models allow to treat the core and halo nucleons 
explicitly, the nuclear volumes of $^6$He and $^8$He can be assumed 
to be composed of an $\alpha$-like core plus 2 and 4 halo neutrons, 
respectively. The nuclear many-body density based on the GG and GO 
parametrizations can be expressed as (\ref{den_ch}). We can further write
\begin{equation}
\rho_{\rm m}(r)=[N_{\rm core}\rho_{\rm core}(r)+N_{\rm halo}
 \rho_{\rm halo}(r)]/A, \label{e7}
\end{equation}
where $\rho_{\rm {core(halo)}}$ are normalized to unity like $\rho_{\rm m}$, and
$N_{\rm core}$ and $N_{\rm halo}$ are the nucleon numbers in the core and halo
volumes, respectively. From Eq.~(\ref{e7}) one obtains the rms matter radius  
of the nucleus as
\begin{eqnarray}
R_{\rm m}=\left[\int r^2\rho_{\rm m}(r)d^3r\right]^{1/2}. \label{e8}
\end{eqnarray}
The core and halo radii ($R_{\rm c}$ and $R_{\rm h}$) are determined by the 
same Eq.~(\ref{e8}) using $\rho_{\rm core}$ and $\rho_{\rm halo}$, respectively. 

\subsection{$\chi^2$-fit procedure for the density parameters}
Each phenomenological density distribution determined above has two free 
parameters (like $R_0$ and $a$ in the SF and WS parametrizations). The aim of the 
present analysis is to find the optimal values of these parameters based on 
the best GMSM description of the experimental data. In the $\chi^2$-fit procedure,
the density parameters are varied independently from each other, and the 
statistical errors as well as the uncertainty in the absolute normalization of the
measured scattering cross sections are taken properly into account \cite{Al02}.

The elastic scattering cross sections at the low and high momentum transfers 
were measured at practically the same energy, and this allows us to combine 
both data sets in the present analysis. Thus, the $\chi^2$ function 
is determined as
\begin{eqnarray}
\chi^2&=&\sum\limits_{j=1}^{N_L}\left[\frac{A_{\rm L}\sigma_{\rm exp}(\theta_j)
-\sigma_{\rm cal}(\theta_j)}{\Delta\sigma_{\rm exp}(\theta_j)}\right]^2 +
\sum\limits_{k = 1}^{N_H}\left[\frac{A_{\rm H}\sigma_{\rm exp}(\theta_k)- 
\sigma_{\rm cal}(\theta_k)}{\Delta \sigma_{\rm exp}(\theta_k)}\right]^2 \nonumber \\
    & & +\left(\frac{A_{\rm L}-1}{\Delta A^{\rm L}_{\rm exp}}\right)^2 + 
		\left(\frac{A_{\rm H}-1}{\Delta A^{\rm H}_{\rm exp}}\right)^2, \label{chi2}
\end{eqnarray}
where $\sigma_{\rm exp}(\theta_j)\equiv
[d\sigma/d\Omega_{\rm c.m.}(\theta_j)]_{\rm exp}$ 
and $\Delta \sigma_{\rm exp}(\theta_j)$ are the experimental differential cross 
sections measured at $\theta_j$ and their statistical errors, and
$\sigma_{\rm cal}(\theta_j)\equiv[d\sigma/d\Omega_{\rm c.m.}(\theta_j)]_{\rm cal}$ 
are the calculated cross sections. $N_L$ and $N_H$ are the number of data 
points measured at low \cite{Neu02} and high momentum transfers \cite{Kis11}, 
respectively. $A_{\rm L}$ and $A_{\rm H}$ are the absolute normalization of the
data points at low and high momentum transfers, and they are treated as free 
parameters in the $\chi^2$-fit, with the estimated uncertainties of the absolute 
calibration $\Delta A^{\rm L}_{\rm exp}\approx 3\%$ \cite{Neu02} and 
$\Delta A^{\rm H}_{\rm exp}\approx 2.4\%$ \cite{Kis11}. 

\section{Results of the GMSM analysis and discussion}
\subsection{The matter radii and matter distributions of $^{6,8}$He}
\begin{figure}[!t]
\includegraphics[angle=0,scale=0.37]{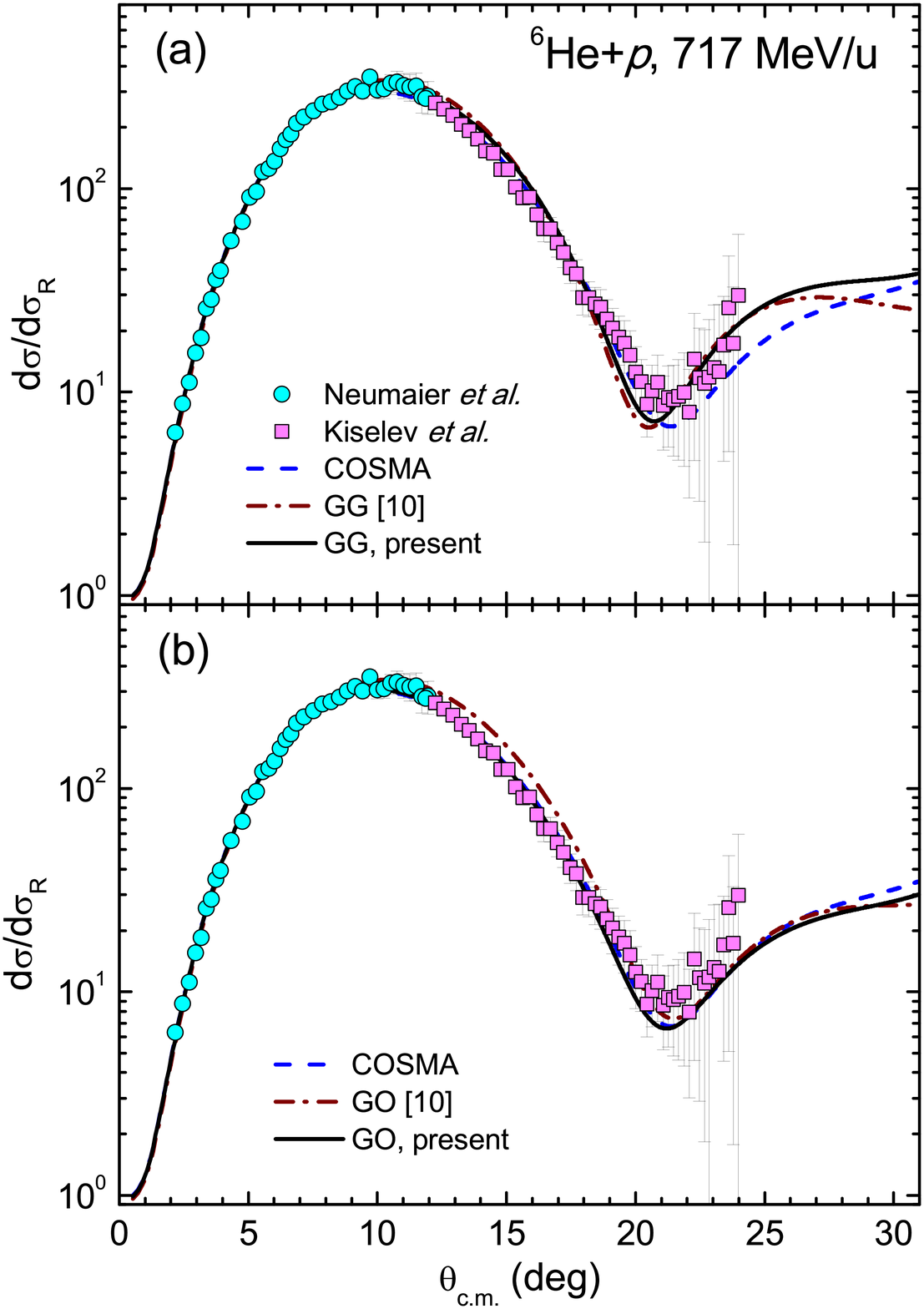}
\caption{(Color online) Elastic \hp6 scattering cross sections (divided by Rutherford cross section) 
obtained with the GMSM calculation (solid curve) using the best-fit parameters of the 
GG (a) and GO (b) models of the nuclear density, in comparison 
with the data measured by Neumaier {\it et al.} \cite{Neu02} and by Kiselev {\it et al.} 
\cite{Kis11} at low and high momentum transfers, respectively. 
The dash-dotted curves were obtained with the best-fit parameters of the GG
and GO density models taken from Ref.~\cite{Al02}, and the dashed curves were 
obtained with the $^6$He density given by the cluster-orbital shell-model 
approximation (COSMA) \cite{Kor97}.}.
\label{He6}
\end{figure}

\begin{figure}[!t]
\hspace{-1cm}
\includegraphics[angle=0,scale=0.38]{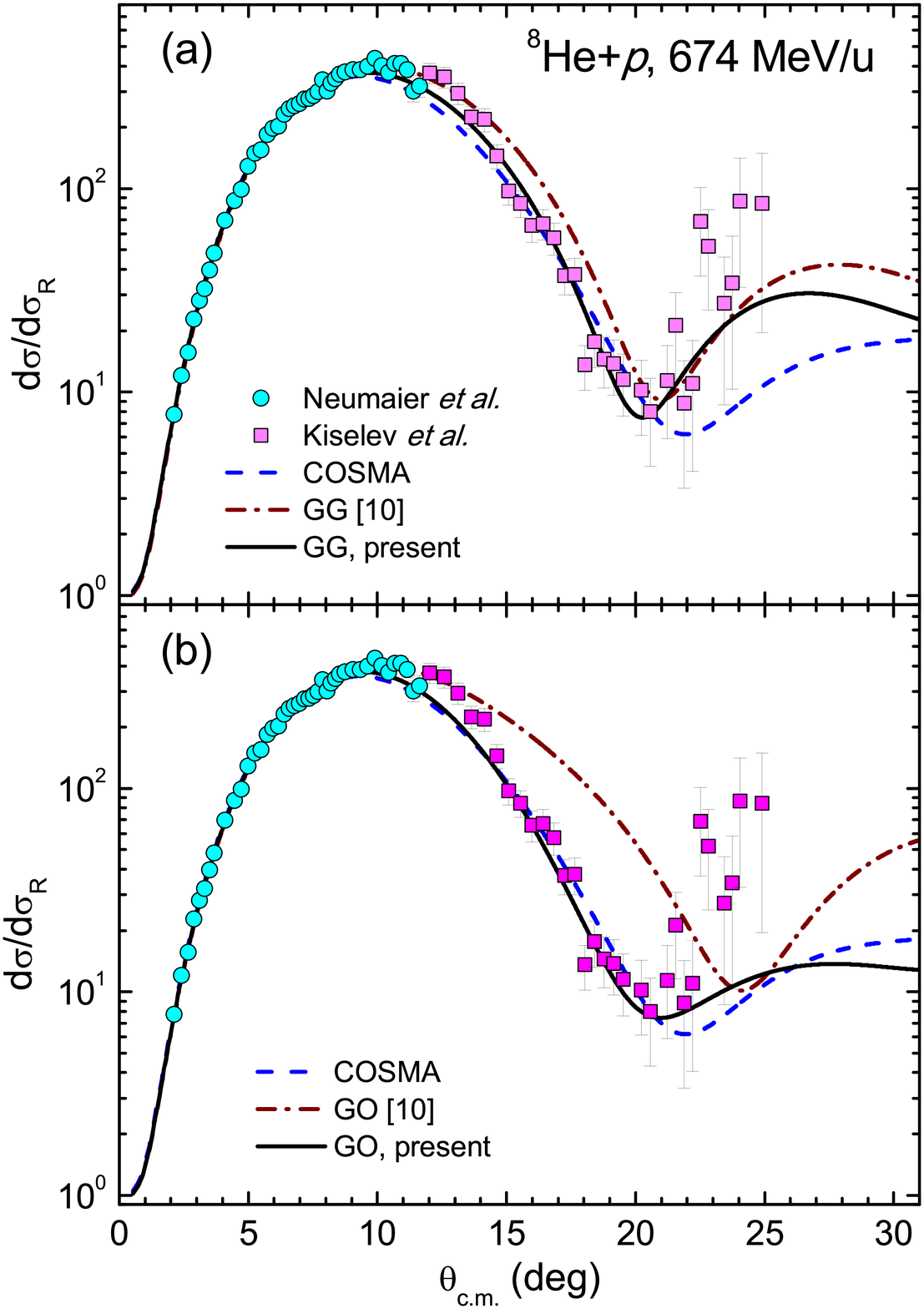}
\caption{(Color online) The same as Fig.~\ref{He6} but for the \ph8 scattering.} \label{He8}
\end{figure}

The $\chi^2$ analysis has been done carefully for each density parametrization 
to obtain the best GMSM description of the elastic $^{6}$He and $^{8}$He 
scattering data measured at the energies of 717 and 674 MeV/u, 
respectively. All the best-fit parameters are presented in Tables \ref{tHe6}
and \ref{tHe8}.     
\begin{table}[!b]
\centering
\caption{The best-fit parameters of the nuclear densities (\ref{SF})-(\ref{GO}) 
obtained from the present GMSM analysis of the combined set of the elastic 
$^{6}$He+$p$ scattering data measured at low \cite{Neu02} and high momentum 
transfer \cite{Kis11}. The relative $\chi^2_{\rm r}$ is per data point, and the
errors are statistical. The neutron radius $R_{\rm n}$ is determined with the
assumption that the proton and core radii are the same, i.e., $R_{\rm p}=R_{\rm c}$. 
The COSMA density (\ref{COSMA}) is parametrized by the same functional as that of the GO 
density model, with the corresponding parameters given in round brackets.}   
\label{tHe6}\vspace{0.5 cm}
\begin{tabular}{|c|c|c|c|c|c|c|c|c|} \hline
density & $A_{\rm L}$ & $A_{\rm H}$ & \multicolumn{2}{c|}{density parameters} & 
$R_{\rm m}$ & $\chi^2_{\rm r}$ & $R_{\rm n}$ & $R_{\rm n}-R_{\rm p}$\\ \cline{4-5}
   &  &  & (fm) & (fm) & (fm) &   & (fm) & (fm)  \\ \hline
GG   &1.04(3)&1.09(4)&$R_{\rm c}$=1.96(4) &$R_{\rm h}$=3.30(12) &2.48(6)&1.41 & 2.71(7) & 0.75(8)\\ \hline   
GO  &1.05(2)&1.04(2)&$R_{\rm c}$=1.90(3) &$R_{\rm h}$=3.26(13) &2.44(5)& 0.88 & 2.67(8) & 0.77(9) \\ \hline   
GH  &1.04(3)&1.09(3)&$R_{\rm m}$=2.45(4)&$\alpha$=0.12(2)         &2.45(4)&1.39 &   & \\ \hline   
SF   &1.05(4)&1.09(3)&$R_0$=1.00(8)           &$a$=0.61(2)        &2.40(5)&1.55 &   &\\ \hline   
WS  &1.04(2)&1.07(3)&$R$=0.99(5)              &$a$=0.63(2)        &2.45(6)&1.00 &   &\\ \hline   
COSMA  &1.00&1.00&$a$=1.55 ($R_{\rm c}$=1.90)&$b$=2.12 ($R_{\rm h}$=3.35) &2.48&1.49 & 2.72  &0.82\\ \hline   
  \end{tabular}
\end{table} 
\begin{table}[!b]
\centering
\caption{The same as table \ref{tHe6} but for the $^{8}$He+$p$ system} \label{tHe8}  
\vspace{0.5 cm}
\begin{tabular}{|c|c|c|c|c|c|c|c|c|} \hline
density & $A_{\rm L}$ & $A_{\rm H}$ & \multicolumn{2}{c|}{density parameters} & 
$R_{\rm m}$ & $\chi^2_{\rm r}$ & $R_{\rm n}$ & $R_{\rm n}-R_{\rm p}$\\ \cline{4-5}
   &  &  & (fm) & (fm) & (fm) &   & (fm) &  (fm)  \\ \hline
GG  &1.00(2)&0.99(6)&$R_{\rm c}$=1.81(6)&$R_{\rm h}$=3.12(13)&2.55(8)&1.35 &2.75(10) & 0.94(12)\\ \hline   
GO  &1.03(2)&0.95(7)&$R_{\rm c}$=1.69(6)&$R_{\rm h}$=2.99(14)&2.43(9)&1.50 &2.63(11) & 0.94(12)\\ \hline   
GH  &1.01(2)&0.98(6)&$R_{\rm m}$=2.50(5)&$\alpha$=0.13(4)      &2.50(5)&1.35 & & \\ \hline
SF   &1.01(2)&0.96(5)&$R_0$=0.66(4)           &a=0.66(2)  &2.51(7)&1.16 & & \\ \hline   
WS &1.01(2)&0.97(5)&$R$=0.80(8)              &a=0.66(2)   &2.51(5)&1.15 & & \\ \hline   
COSMA  &1.00&1.00&$a$=1.38 ($R_{\rm c}$=1.69) &$b$=1.99 ($R_{\rm h}$=3.15) &2.53&2.15 & 2.75  &1.06\\ \hline   
  \end{tabular} 
  \end{table} 
The elastic \phe68 scattering cross sections given by the GMSM calculation 
using the best-fit parameters (see Tables~\ref{tHe6} and \ref{tHe8}) of the 
nuclear matter densities are compared with the data in Figs.~\ref{He6} and \ref{He8}. 
Focusing on the new data points measured at high momentum transfer, one can see 
that the first diffraction maximum in the elastic scattering cross section is now 
fully covered by the data and it turned out that the combined data set allowed 
for an improved determination of the parameters of the density distribution.  
The data and the calculated cross sections divided by the Rutherford 
cross section are presented in Figs.~\ref{He6} and \ref{He8}, and one can see
that the elastic \phe68 scattering at the considered energies is strongly dominated 
by the nuclear scattering, and that allows the fine-tuning of the density inputs 
for the GMSM calculation by the $\chi^2$-fit procedure (\ref{chi2}).  

From a comparison of the best-fit matter radii $R_{\rm m}$ obtained in the present
work for $^{6,8}$He with the results of the earlier GMSM analysis \cite{Al02}
based on the low-momentum data only \cite{Neu02}, we found that the newly obtained 
$R_{\rm m}$ values are slightly larger than those reported in Ref.~\cite{Al02}. 
In terms of the $\chi^2$- fit, the accuracy of the present GMSM analysis is 
about the same as that of Ref.~\cite{Al02}. The nuclear radii obtained in
the present work are also in a sound agreement with the empirical 
matter radii of $^{6,8}$He discussed recently by Tanihata \emph{et al.} in 
Ref.~\cite{Tan13}. The GG and GO density models treat the core and halo parts 
explicitly, and we could determine from our GMSM analysis the neutron skin 
of 0.76(10) and 0.94(13) fm for $^{6}$He and $^{8}$He, respectively. 
Such neutron skins are much thicker than, e.g., the neutron skin 
of around $0.2\sim 0.3$ fm established for the heavy $^{208}$Pb nucleus with 
a large neutron excess, and clearly associated with the halo structure
of the $^{6}$He and $^{8}$He isotopes.   
It is noteworthy that both the GG and GO density models give the best-fit core 
radius for $^{6}$He slightly larger than that for $^{8}$He, and that makes the
difference in the observed neutron skin because the neutron radii are about the
same for the two nuclei. Such an effect was also found in the earlier GMSM
analysis of the elastic \phe68 scattering data taken at low-momentum transfer
\cite{Al02}, and it might be due to different polarizing contributions of the 
valence neutrons to the motion of the $\alpha$-core. Quite complementary
to this discussion are the high-precision laser spectroscopy data  
that yield a charge radius of 2.068(11) fm for $^{6}$He, which is 
significantly larger than the charge radius of 1.93(3) fm obtained
for $^{8}$He \cite{Mue07}. After the standard correction for the finite size 
of the proton \cite{Tan13}, we can obtain the proton radii of 1.925(12) and 1.81(3) 
fm for $^{6}$He and $^{8}$He, respectively, from the laser spectroscopy data. 
Such proton radii are in a good agreement with the core radii of $^{6}$He 
and $^{8}$He given by the present GMSM analysis (see $R_{\rm c}$ values
in Tables~\ref{tHe6} and \ref{tHe8}).    
\begin{figure}[!t]
\centering
\includegraphics[angle=0,scale=0.36]{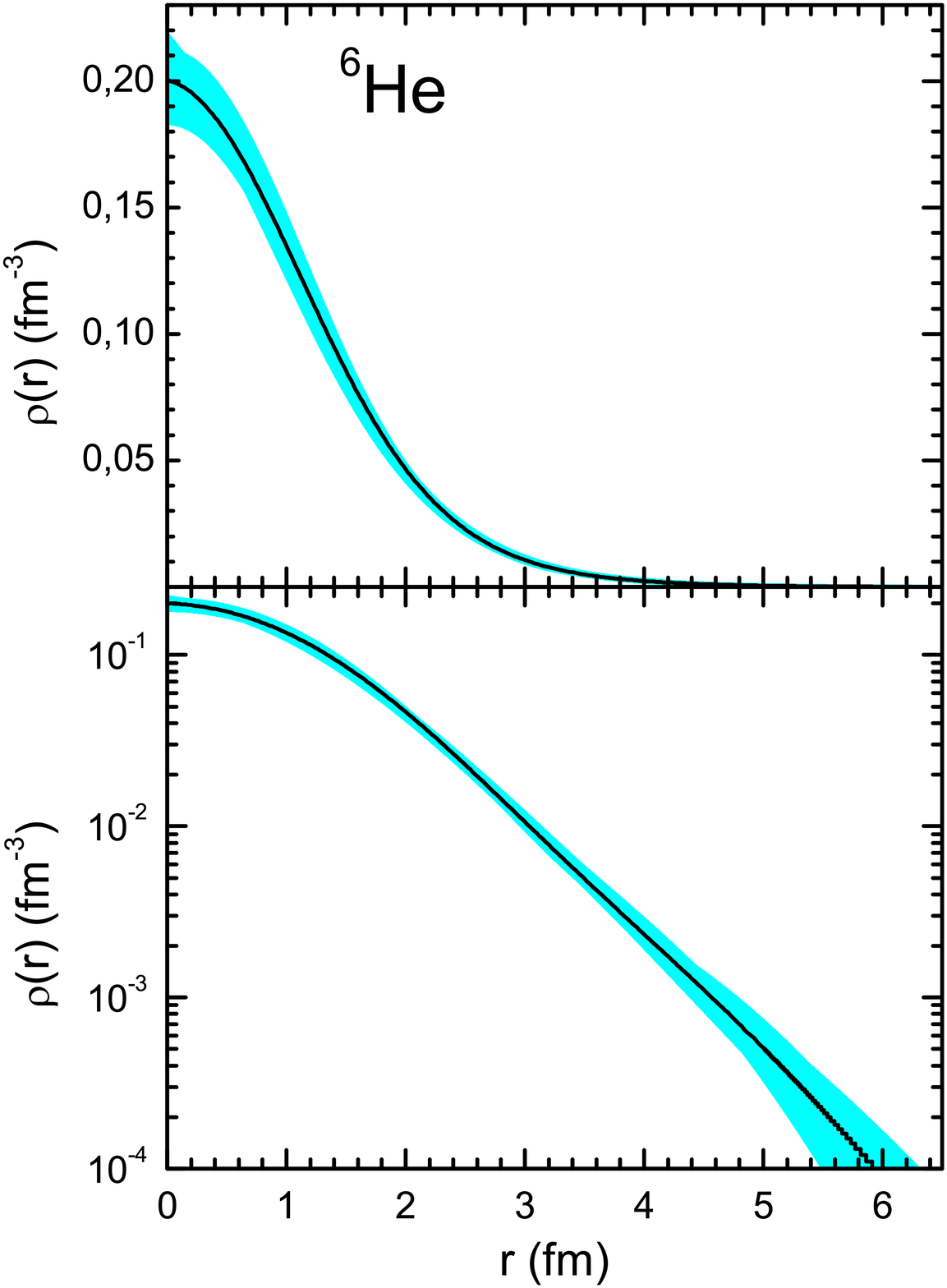} 
\caption{(Color online) The average nuclear matter density distribution of $^6$He  (upper panel) deduced from the 
GMSM fit to the data using the present SF, GH, WS, GG and GO parametrizations, 
with the uncertainty band determined by the statistical errors of the best-fit
parameters of the density models. The same density is plotted in logarithmic scale 
in the lower panel to illustrate the uncertainty at large radii.}  \label{He6_dens}
\end{figure}

To compare with the available results for the microscopic nuclear densities 
predicted by the cluster-orbital shell-model approximation (COSMA) \cite{Kor97}, 
we have also used the COSMA densities as input for the present GMSM calculation,
and the results are presented in Tables~\ref{tHe6} and \ref{tHe8} and Figs.~\ref{He6} 
and \ref{He8}. One can see that the COSMA densities give a good description
of the \hp6 data, but fail to account for the data points taken at angles 
beyond the diffractive minimum for the \ph8 system. Because the newly measured 
data points at large angles allowed us to improve the density parameters of the 
density models, these data are also helpful in fine-tuning the existing 
parameters of the COSMA densities \cite{Kor97}. Namely, from the explicit
expression of the COSMA density
\begin{equation}
\rho_{\rm m}(r)= N_{\rm core}\frac{\exp(-r^2/a^2)}{\pi^{3/2}a^3}+N_{\rm halo}
\frac{2\exp(-r^2/b^2)}{3\pi^{3/2}b^5}r^2, \label{COSMA}
\end{equation}
we find immediately from the best-fit parameters of the GO model in Tables~\ref{tHe6} 
and \ref{tHe8} that the improved $a$ and $b$ parameters of COSMA are 1.55(2) and 
2.06(8) fm, respectively, for $^6$He, and 1.38(5) and 1.89(9) fm for $^8$He. 

\begin{figure}[hbt]
\centering
\includegraphics[angle=0,scale=0.36]{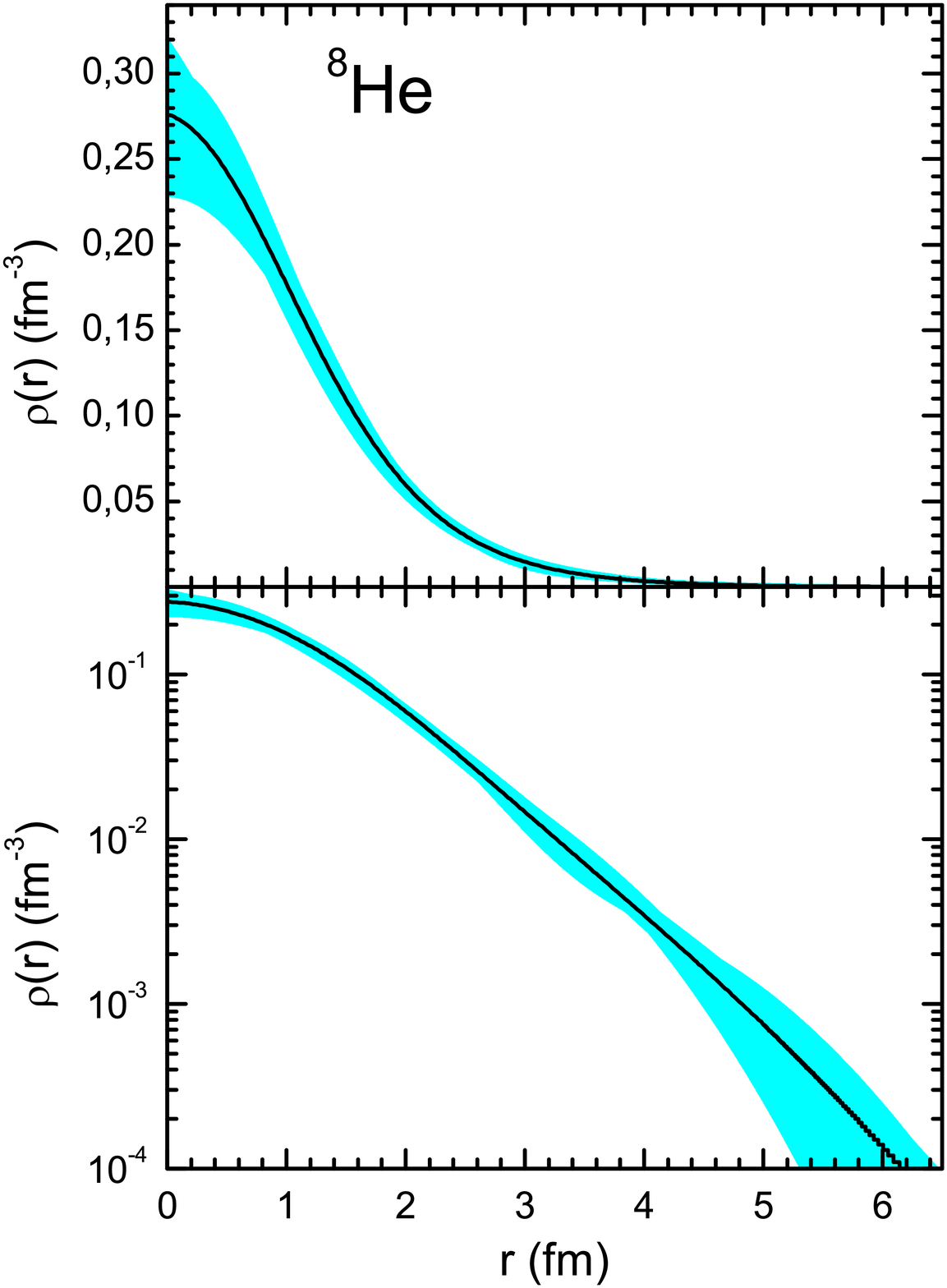} 
\caption{(Color online) The same as Fig.~\ref{He6_dens} but for the matter distribution 
 of $^8$He.} \label{He8_dens}
\end{figure}

With the new parameters of the considered density models given in Tables
\ref{tHe6} and \ref{tHe6}, it is of interest to construct the average 
radial shape of the nuclear matter density distributions for the 
$^6$He and $^8$He isotopes. The radial profiles of the nuclear matter densities 
of the $^6$He and $^8$He isotopes based on the best-fit parameters of 5 density
models are plotted in Figs.~\ref{He6_dens} and \ref{He8_dens}, respectively. 
The errors in tables \ref{tHe6} and \ref{tHe8} are statistical errors 
coming from fitting and data point normalization. The total errors should have included, in 
addition, the contribution from the uncertainties in the $pN$ scattering amplitudes, 
the $t$-scale calibration, and model uncertainty \cite{Al02}. Thus, the final 
averaged nuclear matter radii $R_{\rm m}$ of the $^6$He and $^8$He 
isotopes obtained from the consistent GMSM analysis using the 
five phenomenological parametrizations of nuclear matter density are 
\begin{center}
 $R_{\rm m}=2.44 \pm 0.07$ fm for $^6$He, \\
 $R_{\rm m}=2.50 \pm 0.08$ fm for $^8$He.
\end{center}  
We note further, in connection with the realistic core (or proton) radii of the $^{6}$He 
and $^{8}$He nuclei discussed above, the results of the Glauber few-body calculation \cite{Al98} 
of the elastic \phe68 scattering that gives a very nice description of the data measured 
at low transfer momentum, using the microscopic few-body model that gives the
nuclear matter radii of the $^6$He and $^8$He nuclei $R_{\rm m}\approx 2.50$ and 
2.60 fm, respectively.  These values are somewhat larger than those obtained in 
Ref.~\cite{Al02} (based on the data measured at low transfer momentum) and in the 
present work (based on the complete data set extended to high transfer momentum). 
A likely reason for such a disagreement is the assumption of the rigid $\alpha$-core 
of the fixed radius $R_{\rm c}\approx 1.49$ fm in the few-body calculation, which
is not the case in view of the high-precision laser spectroscopy data \cite{Mue07} 
that give the proton radii of 1.925(12) and 1.81(3) fm for $^{6}$He and $^{8}$He, 
respectively. Such an effect is expected to be due to the different polarizing contributions 
of the valence neutrons to the motion of the $\alpha$-core in these two nuclei 
\cite{Tan13}. It is, therefore, of high interest to have the few-body calculation \cite{Al98}
redone using the quoted experimental values for the $\alpha$-core radius.   

\subsection{Sensitivity of the data to the core and halo parts of the
matter distribution, and to the spin-orbit term}
\begin{figure}[t]
\centering
\includegraphics[angle=0,scale=0.36]{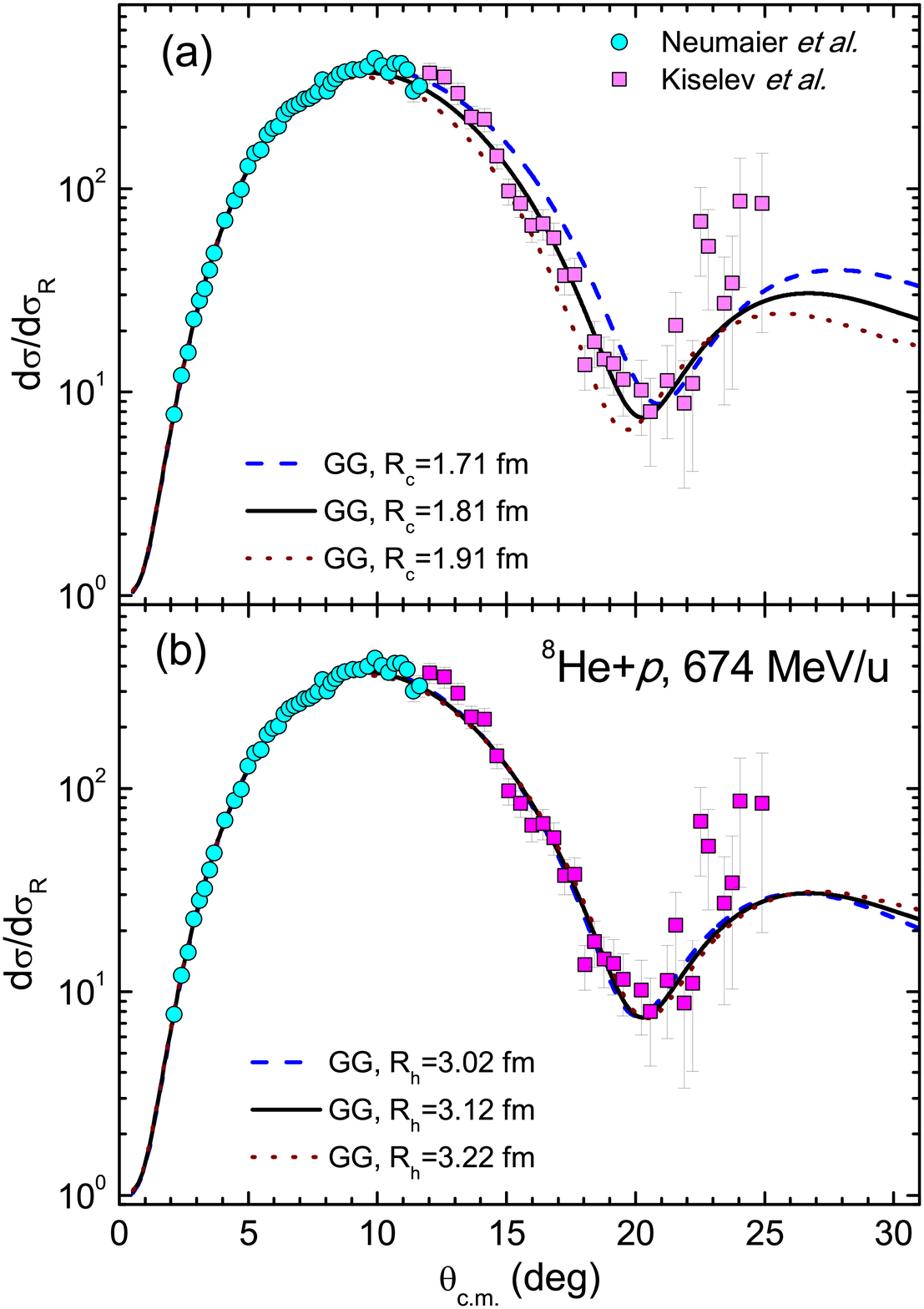} 
\caption{(Color online) The sensitivity of the elastic \ph8 data to the core (a) 
and halo (b) parts of the GG density of $^{8}$He being used in 
the GMSM calculation. See text for more details.} \label{Sens}
\end{figure}
Taking into account the new data taken at high momentum transfer, it is naturally 
to expect that these data points are more sensitive to the inner part of the density 
distribution compared to the sensitivity of the data taken at low momentum 
transfer only. We have made, therefore, some comparisons of the GMSM results 
obtained for the \ph8 case with the halo or core radius of the GG or GO density
model fixed, and the other radius (core or halo) being changed up and down by about 
0.1 fm from the best-fit values given in Table~\ref{tHe8}. From the GMSM results shown 
in the upper panel of Fig.~\ref{Sens} one can see that the data measured at high 
momentum transfer are indeed sensitive to the core part of the density distribution 
of $^{8}$He. A similar variation of the halo radius resulted on a much smaller 
change in the calculated elastic scattering cross section that is hardly 
visible in the logarithmic scale (lower panel of Fig.~\ref{Sens}). Similar 
results were also found for the \hp6 case, and these results confirm that 
the elastic scattering data measured at high momentum transfer are very valuable 
for a precise determination of the core matter density of a halo nucleus. Note
that the GG and GO density parametrizations are defined with an assumption that 
both the $^6$He and $^8$He nuclei have a $\alpha$-like core. The present GMSM 
analysis using the GG and GO density parametrizations has reached a good fit
of the data (see $\chi^2$ values in Tables~\ref{tHe6} and \ref{tHe8}) and
we obtained the following average core and halo radii of the two He isotopes
\begin{center}
 $R_{\rm c}=1.93 \pm 0.06$ fm, $R_{\rm h}=3.28 \pm 0.13$ fm for $^6$He, \\
 $R_{\rm c}=1.75 \pm 0.08$ fm, $R_{\rm h}=3.06 \pm 0.14$ fm for $^8$He.
\end{center}  
As discussed above, the $\alpha$-core radius of $^{6}$He is slightly larger 
than that of $^{8}$He, and the $R_{\rm c}$ values are quite close to the proton 
radii of $^{6}$He and $^{8}$He deduced from the laser spectroscopy data. 
This is a clear indication of the different polarizing contributions of the 
valence neutrons to the motion of the $\alpha$-core in the $^{6,8}$He nuclei.
\begin{figure}[bht]
\centering
\includegraphics[angle=0,scale=0.355]{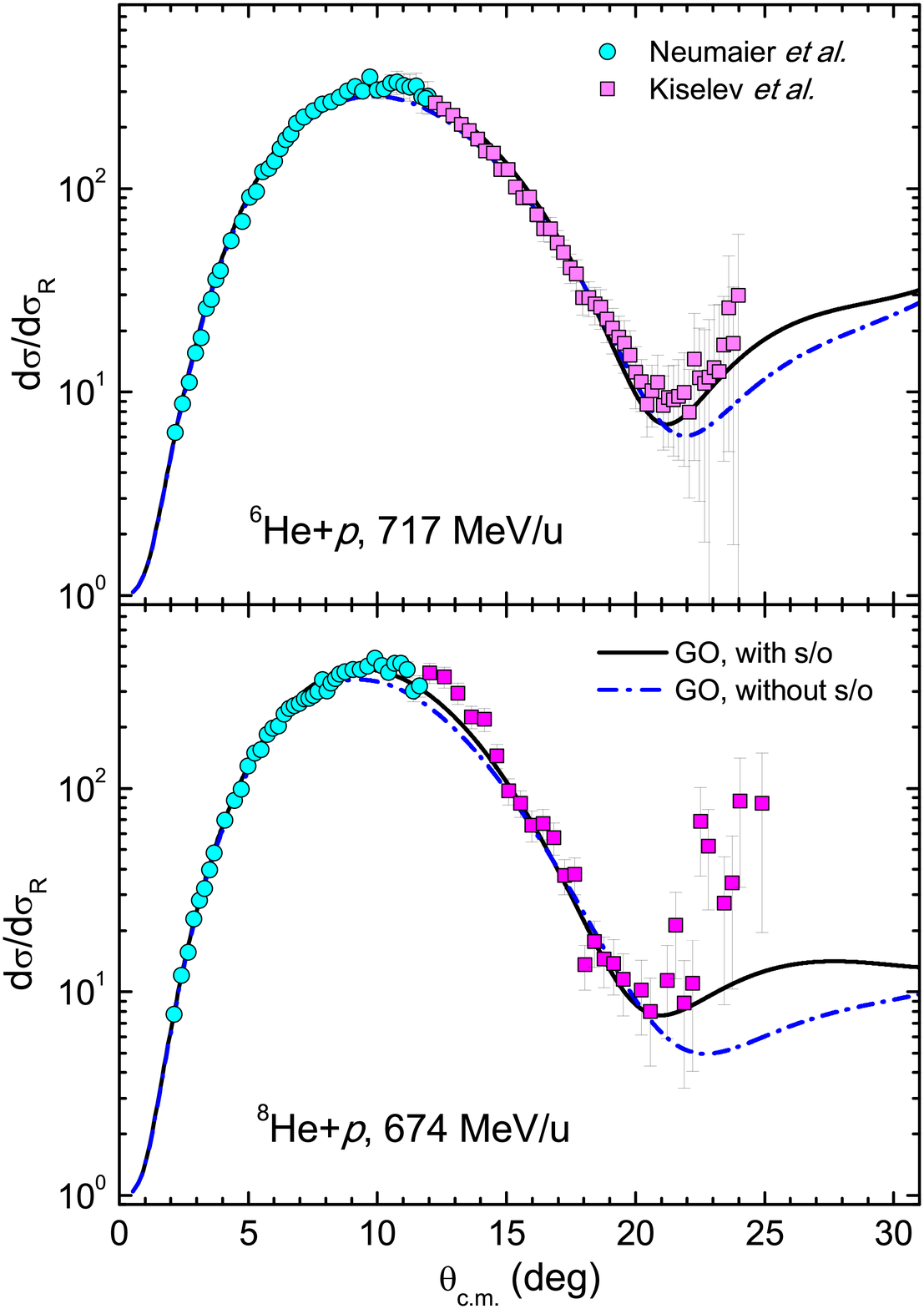} 
\caption{(Color online) Results of the GMSM calculation of the elastic \phe68 scattering
 using the GO density models, with or without the inclusion of the 
 spin-orbit term.} \label{SO_com}
\end{figure}

We note further that the inclusion of the spin-orbit amplitude into the GMSM
calculation is necessary for the analysis of the elastic data measured
at large scattering angles or high momentum transfer. The GMSM results 
plotted in Fig.~\ref{SO_com} show clearly the important contribution 
of the s/o term around the first diffractive minimum as discussed earlier
by Alkhazov \cite{Al78}. The full GMSM calculation with both the central 
and s/o amplitudes included also resulted on slightly larger matter 
radii for $^{6,8}$He nuclei, which are closed to the empirical 
values \cite{Tan13}.

\subsection{Nuclear geometry for the 2-neutron halo in the $^{6}$He nucleus}
\begin{figure}[!t]
\centering
\includegraphics[angle=0,scale=0.8]{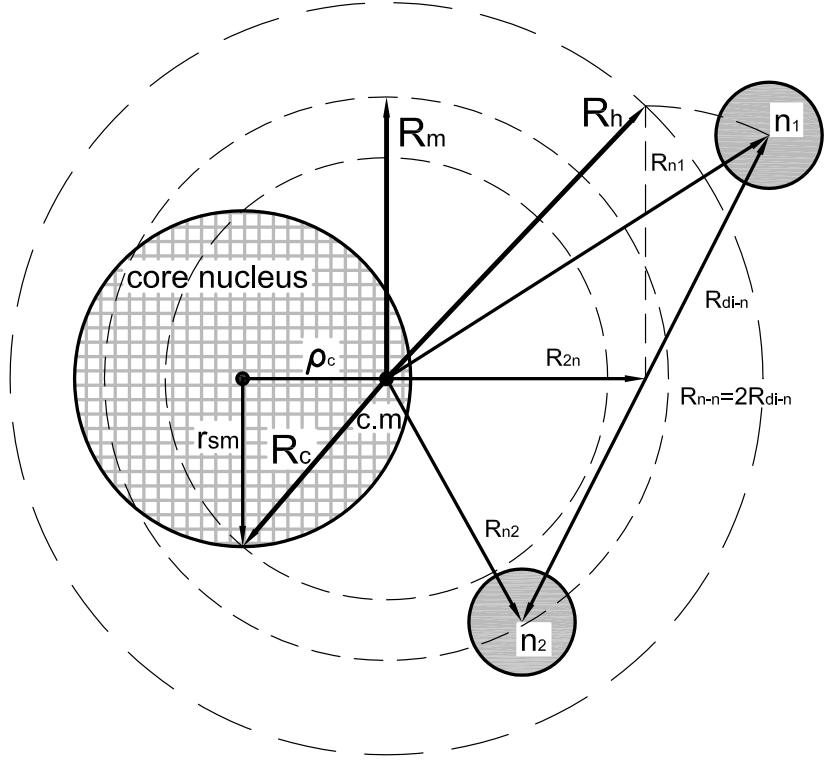} 
\caption{The nuclear geometry for a 2-neutron halo nucleus. See text
for more details} \label{geometry}
\end{figure}

In this section, we apply our GMSM results to the 2-neutron halo geometry 
like that used by Tanihata {\it et al.} in Ref.~\cite{Tan13} for $^{6}$He. 
In this model, the core is assumed to be a free core nucleus that 
moves around the nuclear center of mass, like the 2-neutron halo does. 
As a result, the size of the effective core is bigger than the free 
$\alpha$-particle, and the extended matter distribution is mainly 
determined by the location of the 2-neutron halo. The geometrical model 
of the Borromean $^{6}$He nucleus is shown in Fig.~\ref{geometry},
where the nuclear radii under discussion are defined. 
\begin{table}[bht]
\caption{The radii (in fm) of the geometrical model \cite{Tan13} for the 2-neutron 
halo nucleus $^{6}$He in comparison with the results of the present work.} 
 \vspace{0.5 cm} \label{radii}
\begin{tabular}{|c | c | c | c| } \hline
$^6$He	  & Definition from Ref.~\cite{Tan13} & Present work & Ref.~\cite{Tan13} 	\\ \hline
$R_{\rm m}$		     &$R_{\rm m}$						        & 2.44(7)	& 2.43(3)			\\ \hline
$R_{\rm p }$		     &$R_{\rm c}$						        & 1.93(6) & 1.912(18)		\\ \hline
$R_{\rm h}$		     &$R_{\rm h}$	 					        & 3.28(13) & 3.37(11)	                \\ \hline
$R_{\rm n}$		   &  	 $R_{\rm n}$       & 2.69(9)	& 2.65(4)			\\ \hline
$R_{\rm n}-R_{\rm p}$ & $R_{\rm n}-R_{\rm p}$  & 0.76(10)	& 0.808(47)		\\ \hline
$\rho_{\rm c}$		     &$(R_{\rm c}^2-r_{\rm {sm}}^2)^{1/2}$		& 1.26(7)	&				\\ \hline
$R_{\rm 2n}$		     &$A_{\rm c}/A_{\rm h}\rho_{\rm c}$			& 2.52(13)	& 2.52(5)			\\ \hline
$R_{\rm c-2n}$	     &$\rho_{\rm c}+R_{\rm 2n}$			        & 3.79(14)& 3.84(6)			\\ \hline
$R_{\rm di-n}$		     &$(R_{\rm h}^2-R_{\rm 2n}^2)^{1/2}$		& 2.09(25)	&						\\ \hline
$R_{\rm n-n}$		     &$2R_{\rm di-n}$				                & 4.19(49)& 3.93(25)		\\ \hline
$\bm R_{\rm n1}.\bm R_{\rm n2}$ & $(A_{\rm c}^2\rho_{\rm c}^2-R_{\rm n-n}^2)/4$&1.99(119)&	2.70(97)		\\ \hline
   \end{tabular}
\end{table} 

Because the core is an $\alpha$-cluster, the matter, proton and neutron radii 
of the core nucleus can be assumed equal  
$r_{\rm {sm}}=r_{\rm {sp}}=r_{\rm {sn}}=1.46$ fm \cite{Tan13}.
Using these values and $R_{\rm m}$, $R_{\rm c}$, $R_{\rm h}$, $R_{\rm n}$ radii 
given by the present GMSM analysis with the GG and GO density models, 
the radii of the geometry shown in Fig.~\ref{geometry} can be determined 
\cite{Tan13} as.  
\begin{itemize}
	\item The distance $\rho_{\rm c}$ between the nuclear center of mass and 
  the core center is 
\begin{equation}
\rho_{\rm c}=\sqrt{R_{\rm c}^2-r_{\rm {sm}}^2}. \label{geob}
\end{equation}
\end{itemize}
\begin{itemize}
	\item The vector $R_{\rm 2n}$ joining the nuclear center of mass and the midpoint 
of the line connecting the two halo neutrons is determined from the balancing condition
\begin{equation}
 A_{\rm c}\rho_{\rm c}=A_{\rm h}R_{\rm 2n}, \ \mbox{where}\ A_{\rm c}=4,\ A_{\rm h}=2.
\end{equation}
\end{itemize}
\begin{itemize}
	\item The distance $R_{\rm c-2n}$ from the core center to the two halo neutrons is 
\begin{equation}
 R_{\rm c-2n}=\rho_{\rm c}+R_{\rm 2n}.
\end{equation} 
\end{itemize}
\begin{itemize}
	\item The distance $R_{\rm n-n}$ between the two halo neutrons is given by 
\begin{equation}
 R_{\rm n-n}=2R_{\rm di-n},\ \mbox{where}\ R_{\rm h}^2=R_{\rm 2n}^2+R_{\rm di-n}^2.
\end{equation}
\end{itemize}
\begin{itemize}
	\item The radial correlation of the two halo neutrons is determined as
\begin{equation}
 \bm R_{\rm n1}.\bm R_{\rm n2}=(A_{\rm c}^2\rho_{\rm c}^2-R_{\rm n-n}^2)/4. \label{geoe}
\end{equation}
\end{itemize}

The results obtained for the considered geometrical model of $^{6}$He are 
summarized in Table~\ref{radii}, and one can see a good agreement of our results 
with those determined in Ref.~\cite{Tan13}. 
Despite the simplicity, the considered geometrical model gives a good illustration 
of the core movement inside a 2-neutron halo nucleus, which can be estimated by 
the difference between the core matter radius and that of the free core nucleus. 
For the $\alpha$-core this is just the difference between the proton radius 
of the free $\alpha$-particle and that of the considered halo nucleus because 
protons are distributed in the $\alpha$-core only. In a similar manner, one 
might suggest the geometry for $^8$He, but in this case 4 halo 
neutrons are distributed uniformly against the $\alpha$-core, and the polarizing 
contributions of the valence neutrons to the motion of the $\alpha$-core should  
be weaker than that found in the case of $^6$He. A direct consequence is a smaller core
radius of $^8$He compared to that of $^6$He as found in the present GMSM analysis. It is noted that the $^8$He geometry can be considered to have 2 halo neutrons \cite{Chu05}.  In this case, the same procedure is applied to determine the nuclear radii as discussed above but now the compact core is $^6$He.
\section{Summary}
The detailed GMSM analysis of the latest experimental data of the elastic \phe68
scattering at 717 and 674 MeV/u has been performed. Based on the new data points  
measured up to the first diffractive minimum, the nuclear radii as well as the
radial shape of the matter distribution of these helium halo nuclei have been
determined, and the results are in a sound agreement with the
recent systematics of these quantities given in Ref.~\cite{Tan13}.  

The sensitivity of the new data points taken at large momentum transfer to the
core radius of the $^{6,8}$He nuclei as well as to the spin-orbital 
term in the GMSM calculation was demonstrated. The combined data set
taken at both low and high momentum transfer were used to fine-tune the parameters
of the nuclear densities of $^{6,8}$He based on the cluster-orbital shell-model 
approximation \cite{Kor97}.    

The core and halo radii obtained from the present GMSM analysis were used in a 
geometrical model suggested for the Borromean nucleus $^{6}$He \cite{Tan13} to 
determine various size parameters of this nucleus, and the results agree with 
those obtained in Ref.~\cite{Tan13}. The enhancement of the $\alpha$-core radius 
of $^{6}$He compared to that of $^{8}$He found in the present GMSM analysis can be  
qualitatively understood in that simple geometrical picture. 

\section*{Acknowledgments}
The present research has been supported, in part, by the National Foundation 
for Science and Technology Development (NAFOSTED project No.103.04-2014.76)
and by the Ministry of Science and Technology of Vietnam (project No.105/2013/HD$-$NDT). 
We are grateful to Prof. G.D. Alkhazov and authors of Ref. \cite{Al02} for providing us 
with the earlier version of the GMSM code, based on which we have developed 
the present version that includes the spin-orbital amplitude.

\end{document}